\documentclass[11pt]{article}
\usepackage{authblk}

\usepackage{amsmath,graphicx}
\usepackage[round,numbers, sort&compress]{natbib}
\DeclareMathOperator*{\argmin}{\arg\!\min}

\makeatletter
\def\tagform@#1{\maketag@@@{[#1]\@@italiccorr}}
\makeatother
\usepackage{setspace}
\usepackage[margin=1in]{geometry}
\usepackage{amssymb}
\usepackage{epstopdf}
\usepackage{subfigure}
\usepackage{amsthm}
\newcommand*{\permcomb}[4][0mu]{{{}^{#3}\mkern#1#2_{#4}}}
\newcommand*{\comb}[1][-1mu]{\permcomb[#1]{C}}
\usepackage{algorithm}
\usepackage{algorithmic}
\let\Algorithm\algorithm
\renewcommand\algorithm[1][]{\Algorithm[#1]\setstretch{1.4}}

\title{\uppercase{ \bf \large \center Multi-shot multi-channel diffusion data recovery using structured low-rank matrix completion}}

\author[1]{\it \small Merry Mani}
\author[2]{\it \small Mathews Jacob}
\author[3]{\it \small Douglas Kelley}
\author[4]{\it \small Vincent Magnotta}
\affil[1]{\it \small Department of Psychiatry, University of Iowa, Iowa City, Iowa}
\affil[2]{\it \small Department of Electrical and Computer Engineering, University of Iowa, Iowa City, Iowa}
\affil[3]{\it \small Global Applied Science Laboratory, GE Healthcare}
\affil[4]{\it \small Department of Radiology, University of Iowa, Iowa City, Iowa}

\begin{document}
\maketitle

\vspace{50mm}
Correspondence to :\\
Mathews Jacob\\
3314 Seamans Center for the Engineering Arts and Sciences\\
Iowa City, Iowa, 52242\\
email: mathews-jacob@uiowa.edu \\
phone number: (319) 335-6420.\\
\\
Word count : 4946\\
Figures+ tables count : 9\\
\\
Running title: Annihilating filter k-space formulation for multi-shot DWI recovery \\

\newpage
{ \bf Abstract}

Purpose: To introduce a novel method for the recovery of multi-shot diffusion weighted (MS-DW) images from echo-planar imaging (EPI) acquisitions.\\
Methods: Current EPI-based MS-DW reconstruction methods rely on the explicit estimation of the motion-induced phase maps to recover the unaliased images. In the new formulation, the k-space data of the unaliased DWI is recovered using a structured low-rank matrix completion scheme, which does not require explicit estimation of the phase maps. The structured matrix is obtained as the lifting of the multi-shot data. The smooth phase-modulations between shots manifest as null-space vectors of this matrix, which implies that the structured matrix is low-rank. The missing entries of the structured matrix are filled in using a nuclear-norm minimization algorithm subject to the data-consistency. The formulation enables the natural introduction of smoothness regularization, thus enabling implicit motion-compensated recovery of fully-sampled as well as under-sampled MS-DW data. \\
Results: Our experiments on in-vivo data show effective removal of the ghosting artifacts arising from inter-shot motion in MS-DW data using the proposed method.  The performance is comparable and better in certain cases than conventional phase-based methods.\\
Conclusion: The proposed method can achieve effective unaliasing of fully/under-sampled MS-DW images without using explicit phase estimates.\\

\vspace{2em}
Keywords: structured low rank, annihilating filter, multi-shot diffusion, calibration-less, motion compensation, regularized recovery .

\newpage

{\Large Introduction}\\ 
Magnetic resonance diffusion-weighted imaging (DWI) is a unique non-invasive tool used to study the micro-architecture of tissues by mapping the diffusion of water molecules in the tissue \cite{JBasser2000, LeBihan1986}. It is widely used in the clinical diagnosis of acute stroke, tumors and brain abscesses \cite{Moseley1990b,Lai2002,Kim1998,Ebisu1996,Skare2007} and has also become the primary neuroscience research tool for studying white matter connections \cite{Johansen-Berg2009}. Single-shot EPI (ss-EPI) acquisition coupled with parallel imaging is currently the preferred method for DWI acquisition, primarily due to its immunity to bulk motion and its short acquisition time \cite{Jaermann2004,Bhagat2007,Schoenberg1991}.  However, the long readouts of the EPI sequences lead to signal loss and blurring while intrinsically low bandwidth along the phase-encoding direction results in distortions along the phase encoding direction\cite{LeBihan2006}. This is a fundamental limitation, which makes ss-EPI less preferable for many applications. For example, high resolution diffusion imaging requires long EPI readout durations and results in considerable geometric distortion and blurring in the resulting images. Another example where the long echo-train length of EPI is undesirable is diffusion imaging at ultra-high field strength (UHFS) where the T2 relaxation times are shortened \cite{Uludag2009}. In such situations, the echo-train length, even for a moderate imaging resolution, is long enough to cause loss of SNR. This limits the ability of ss-EPI-based DWI to leverage the advantages of UHFS MR imaging \cite{Heidemann2012,Lundell2014,Jeong2013}. Acquisitions with shorter readouts would therefore significantly enhance DWI especially when applied at UHFS. \\

Multi-shot sensitivity encoded diffusion weighted imaging (MS-DWI) holds great potential for enabling high spatial resolution diffusion imaging\cite{Skare2007}. The technique can also achieve shorter echo times (TE) to enhance studies at higher field strengths as well as to examine structures that are proximal to inhomogeneous fields or imaging near metal implants. However, MS-DWI has certain limitations. The main problem is the sensitivity of MS-DWI reconstructions to motion resulting from the use of large diffusion gradients, which are designed to encode diffusion induced microscopic motion of the water molecules. Sub-pixel physiologic motion (e.g due to cardiac pulsations, respiratory motion) during the diffusion encoding gradients result in the image being corrupted by smooth spatially varying motion-induced phase \cite{Anderson1994,Bammer2003}. While this phase term does not pose a challenge in single shot imaging, the difference in the phase distortions between aliased multi-shot images manifest as ghosting artifacts in the image obtained by combining the shots. Moreover, the imaging time of MS-DWI increases by a scale factor proportional to the number of shots.   \\

Special reconstruction schemes are required for the MS-DWI to eliminate the shot-to-shot ghosting artifacts resulting from bulk motion as well as physiological motion. Such schemes generally involve a multi-step process, where the shot-to-shot phase variations are first estimated and then applied during image reconstruction. Methods to estimate the phase falls into two categories: (i) methods that rely on separate navigator scans \cite{DeCrespigny1995,Ulug1995,Butts1996, Jeong2013} and (ii) methods that estimates phase from the data itself \cite{Liu2004,Nunes2005,Pipe2002, Uecker2009a, Chen2013,Chu2015,Guhaniyogi2015,Chang2015}. Since phase navigation methods require additional scan time and do not accurately capture the bulk motion during the diffusion scans, estimating phase from the data itself is the more attractive option. The data-based techniques work extremely well for non-Cartesian self-navigated trajectories such as SNAILS, PROPELLER etc where a low-resolution phase map can be obtained from the fully sampled k-space data of each shot \cite{Liu2004,Nunes2005,Pipe2002}. The same strategy has been employed for Cartesian MS-DW acquisition, even in the absence of fully sampled center k-space data\cite{Chen2013}. The estimate of the phase map in this case is obtained by employing highly regularized reconstructions of the individual shot data. However, it should be noted that an {\it{estimate of motion-induced phase maps is a prerequisite for existing unaliasing methods}}. In many cases, a good estimate of motion induced phase is hard to compute. For example, if  multi-shot acquisitions are under-sampled using either parallel imaging or compressed sensing, then the estimate of phase from the data will be unreliable. As a result, effective correction of motion induced ghosting artifacts cannot be accomplished.\\

The aim of this work is to introduce a novel reconstruction for MS-DWI data {\it{that does not rely on the explicit estimation of motion induced phase estimates}} to correct the ghosting artifacts resulting from inter-shot motion. This work exploits the recent advances in multichannel MR image recovery, which utilizes the annihilation relations between the sensitivity weighted images or their Fourier samples \cite{Morrison2007,Shin2014,Haldar2014,Haldar2015}. We adopt the above scheme for the recovery of MS-DWI data by constructing a structured matrix by the so-called "matrix-lifting" of the k-space samples of shots (the specific matrix-lifting is illustrated in  figures 2 and 3), which is low-rank due to the annihilation relations. This property enables us to fill in the missing entries using a structured low-rank matrix completion approach. We also exploit the recent advances in reformulating smoothness regularization as structured low-rank problem, where a similar lifting strategy is adopted \cite{Ongie2015a,Ongie2015,Jin2015,Lee2016,Jin2015a}. We introduce a novel lifting scheme that combines the above structures, which enables the recovery of the motion compensated multishot images from noisy and undersampled diffusion weighted multishot MRI data. We demonstrate the results of the proposed method on in-vivo data collected at 7T.\\

{ \Large Theory}

{  \bf k-Space data matrix structure of a multi-shot DWI acquisition }
\vspace{2mm}\\
In an $N_s$-shot EPI-based sensitivity-encoded diffusion acquisition for sampling an $N_1 \times N_2$ imaging matrix, the readout is shortened by collecting only $N_2/N_s$ phase encoding lines during every acquisition. The acquisition is repeated $N_s$ times, each time sequentially collecting the next set of  $N_2/N_s$ phase-encoding lines (see figure 1(a) for a cartoon of a 4-shot acquisition). In figure 1(b), we represent the 4-shot acquisition using four k-space data matrices concatenated along the shot dimension. The acquired k-space samples are marked using solid circles and the unacquired k-space samples are marked using hollow circles. Note that if phase differences due to inter-shot motion are absent, then we can fill a single k-space data matrix with the sampled points occupying the appropriate positions in the data matrix. In fact, this method is often used in the recovery of the non-diffusion weighted images collected as part of the MS-DWI acquisition. During a diffusion-weighted acquisition, the diffusion sensitizing gradients add extra phase to the moving spins and the acquired k-space data for each shot will have a different phase due to the shot-to-shot inherent sub-pixel motion of the imaging sample. Hence, the acquired k-space data from the separate shots cannot be combined directly; instead they are stacked into separate matrices. Our aim is to recover the unaliased k-space data samples in each of the four k-space data matrices based on the samples that we collected.\\

\vspace{-2mm}
{ \bf Annihilating filter formulation for multi-shot diffusion data}
\vspace{2mm}\\
 Liu et al showed that by using an encoding function that combines the coil sensitivity with the individual shot phase information, the unaliased image could be recovered \cite{Liu2005} using an iterative sensitivity encoded reconstruction algorithm \cite{PPruessmann2001}. The work of Lui et al \cite{Liu2005} combined with the recent null-space based MR image reconstruction methods \cite{Griswold2002,Lustig2010,Zhang2011,Uecker2014} suggests the possibility of using composite sensitivities as a null-space constraint for the recovery of the MS-DWI data. Since null-space methods can be tied to the notion of annihilating filters in frequency domain \cite{Ongie2015}, it is likely that shift-invariant k-space filters corresponding to the composite sensitivities or the motion-induced phase maps exist. Utilizing such filters, the above reconstruction problem can be posed as a structured low rank matrix completion problem.\\

\vspace{-2mm}
{ \bf Structured low-rank property of multi-shot diffusion data  }
\vspace{2mm}\\
In this section, we revisit the relationship between the diffusion weighted data from the different shots of a given direction by rewriting them in the frequency domain. Assuming that the motion-induced phase maps are smooth functions, these can be represented in k-space as shift-invariant filters of finite support \cite{Haldar2014}. The annihilation relations that we obtain based on the k-space filters lead to a low-rank recovery for the MS-DWI data that do not require motion-induced phase estimates as derived in the sections below. \\

Let the complex diffusion weighted image of a given diffusion direction be denoted as $\rho({\bf x})$, where $\bf{x}$ represents spatial co-ordinates. Then, due to the inter-shot motion, the measured DWI from the $l^{th}$ shot will have a phase term \cite{Liu2005} that will be different from the measured DWI from the $i^{th}$ shot, leading us to write 
\begin{equation}
	\label{first_eq1}
	 m_{l}({\bf x})=\rho({\bf x})~\phi_{l}({\bf x}); \forall \bf x.
	 \end{equation} and 
	 \begin{equation}
	\label{first_eq2}
	 m_{i}({\bf x})=\rho({\bf x})~\phi_{i}({\bf x}); \forall \bf x.
	 \end{equation}
Here, we assume that the phase$~\phi({\bf x})$ is also a complex quantity such that $|~\phi({\bf x})|=1; \forall \bf x$.  
 Multiplying Eq. [\ref{first_eq1}]  by  $\phi_{i}({\bf x})$ and Eq. [\ref{first_eq2}] by $\phi_{l}({\bf x})$, we can write
	\begin{equation}
	\label{first_eq}
	 m_{l}({\bf x})\phi_{i}({\bf x})-m_{i}({\bf x})~\phi_{l}({\bf x}) = 0; \forall \bf x.
	 \end{equation}
Taking the Fourier transform on both sides of \eqref{first_eq}, we obtain
	\begin{equation}
	\label{first_eq3}
	 \widehat{m_{l}}[{\bf k}]*\widehat{\phi_{i}}[{\bf k}]-\widehat{m_{i}}[{\bf k}]*\widehat{\phi_{l}}[{\bf k}] = 0; \forall \bf k,
	 \end{equation}
	 where $\widehat{m_{l}}[\bf k]$ and $\widehat{\phi_{l}}[\bf k]$ denote the Fourier coefficients of ${m_{l}}(\bf x)$ and ${\phi_{l}}(\bf x)$, respectively for $l=1,.. N_s$. 
	We observe that $\phi_{l}({\bf x})$ are smooth functions and hence can be reliably approximated as a bandlimited functions. This implies that $\widehat{\phi_{l}}({\bf k}); l=1,.. N_s$ are support limited in k-space. Assuming the support of $\widehat{\phi_{l}}(\bf{k})$ to be $r \times r$, the convolution with this filter can be implemented as multiplication using block Hankel matrices:
	\begin{equation}
	\label{sec_eq}
	{\mathbf{H}}(\widehat{{m}_{l}})\cdot \widehat{\boldsymbol\phi}_{i}-{\mathbf {H}}(\widehat{{m}_{i}})\cdot \widehat{\boldsymbol\phi}_{l} = \bf 0. 
	\end{equation}
Here, ${\bf{H}}(\widehat{m_l}))$ is a block Hankel matrix of size $(N_1-r+1)(N_2-r+1)\times r^2 $ generated from the $N_1 \times N_2$ Fourier samples of $\widehat {m_l}[\bf{k}]$. Each of the rows of ${\bf{H}}(\widehat{m_l})$ are vectorized versions of $r\times r$ rectangular k-space neighborhoods of $\widehat {m_l}[\bf k]$. Figure \ref{fig:fig2} shows the structure of the block-Hankel matrix. In practice, this mapping can be achieved using a sliding window operation as indicated in Figure \ref{fig:fig2}. Similarly, $\widehat{\boldsymbol \phi}_{l}$ is a vector of size $r^2\times 1$, which is the vectorized version of the $r \times r$ filter $\phi_l$. Thus, ${\mathbf{H}}(\widehat{{m}_{l}})\cdot \widehat{\boldsymbol\phi}_{i}$ provides a vector that corresponds to the convolution $ \widehat{m_{l}}[{\bf k}]*\widehat{\phi_{i}}[{\bf k}]$ within the rectangular region $(N_1-r+1) \times (N_2-r+1)\times r^2$. Note that this region corresponds to the valid convolutions between the $r\times r$ filter $\widehat{\phi_{l}}$ and the $N_1\times N_2$ samples of $ \widehat{m_{l}}[{\bf k}]$. 
Since, this relation holds true for all the shots, we can combine the annihilation relations in the matrix form as:\\
\begin{equation}
	\label{equation 3}
	\underbrace{\begin{bmatrix} {\bf{H}}(\widehat{m_1}) & {\bf{H}}(\widehat{m_2}) & ... & {\bf{H}}(\widehat{m_{N_s}}) \end{bmatrix}}_{{{\bf{H}}}_1(\bf\widehat{m})}\underbrace{\left[ \begin{array}{c} \widehat{\boldsymbol\phi_2} \\ -\widehat{\boldsymbol\phi_1}\\ 0 \\ 0\\  \vdots \\ 0 \end{array} \begin{array}{c} 0\\
 \widehat{\boldsymbol\phi_3} \\ -\widehat{\boldsymbol\phi_2}\\ 0 \\ \vdots \\ 0 \\ \end{array}\begin{array}{c} \cdots \end{array}
\begin{array}{c} 0\\0 \\ \vdots \\ 0 \\ \widehat {\boldsymbol\phi_{N_s}}\\
-\widehat {\boldsymbol\phi_{N_s-1}}  \end{array}\begin{array}{c}  \widehat{\boldsymbol\phi_3} \\0\\-\widehat{\boldsymbol\phi_1} \\ 0\\  \vdots \\ 0 \end{array}  \begin{array}{c} \cdots \end{array} \right]}_{\bf{\hat{\Phi}}}= \begin{bmatrix}0 & 0 & ...& 0 & 0 ... \end{bmatrix}. 		\end{equation}
The above relation can be compactly expressed as ${{\bf{H}}}_1(\bf \widehat{m})~\bf{\hat{\Phi}} = 0$.

For an $N_s$-shot acquisition, there will be ${N_s}\choose{2}$ columns in ${\bf{\widehat{\Phi}}}$, which implies that the block-Hankel matrix $\bf{{\bf{H}}}_1(\widehat{m})$ has a null-space of dimension $\ge N_s$. Equivalently, we can say that $\bf{{\bf{H}}}_1(\widehat{m})$ is of low-rank.  Thus, even if we don't have an estimate of $\bf{\widehat{\Phi}}$, we can impose a low-rank penalty on the matrix $\bf{{\bf{H}}}_1(\widehat{m})$. \\

{ \bf MUSSELS: MUlti-Shot Sensitivity Encoded diffusion data recovery using Structured Low rank matrix completion }
\vspace{2mm}\\
In a typical MS-DWI acquisition, the diffusion weighted data are collected as a multi-channel sensitivity-encoded acquisition. The sensitivity-encoding provides additional constraints to recover the unaliased DWIs from a multi-shot acquisition. Two sensitivity-encoded formulations are possible for the recovery of the MS-DWI data: (i) using a regular SENSE formulation and (ii) using annihilating filter relations derived from sensitivity encoding. We use the SENSE-based formulation in this work because of the ease of such implementations. For the sake of completion, the second formulation is provided in the appendix. We assume the coil sensitivities to be estimated from the non diffusion-weighted image are collected as part of the acquisition, which is typically the case. The knowledge of the coil sensitivity information allows us to formulate the recovery of the multi-shot k-space data as follows:
\begin{equation}
\label{use Hankel}
\widehat{\tilde {\bf m}}=  \argmin_{{\bf{\widehat{m}}}} \underbrace{||{\mathcal A}\left({\bf{\widehat{m}}}\right)-{\bf{\widehat{y}}}||^2_{\ell_2}}_{\mbox{data consistency}} + \lambda \underbrace{||\bf{\bf{H_1}}({\bf{\widehat{{m}}}})||_*}_{\mbox{regularization}} .
\end{equation}
Here, $\bf{\widehat y}$ is the measured multi-channel multi-shot k-space data of dimension $N_1\times N_2/N_s \times N_c\times N_{s}$. The first term imposes data consistency to the measured k-space data using a regular SENSE formulation. The operator $\bf{{\cal{A}}}$ represents the concatenation of the following operations: $\bf{{\cal{M}}}\circ\bf{{\cal{F}}}\circ\bf{{\cal{S}}}\circ\bf{{\cal{F}}^{-1}}$ where $\bf{{\cal{F}}}$ and $\bf{{\cal{F}}^{-1}}$ represent the Fourier transform and the inverse Fourier transform operations respectively, $\bf{{\cal{S}}}$ represents multiplication by coil sensitivities, and $\bf{{\cal{M}}}$ represents multiplication by the k-space sampling mask corresponding to each shot. The second term in \eqref{use Hankel} is the nuclear norm of the block-Hankel matrix of the shots, which is the convex relaxation for the rank penalty.  $ \lambda$  is a regularization parameter. We propose to solve  [\ref{use Hankel}] using an augmented Lagrangian minimization scheme. We will refer to \eqref{use Hankel} as MUSSELS recovery scheme, which can recover the multi-shot k-space data corresponding to the unaliased DWI data. Once the multi-shot k-space data is recovered using MUSSELS, then we can recover the magnitude DWI by taking the inverse Fourier transform and doing a sum-of-squares.
\begin{equation}
\begin{split}
	\label{}
	\sum_{l=1}^{N_s} \lvert m_l({\bf x}) \lvert^2=\lvert \rho({\bf x}) \lvert^2 \sum_{l=1}^{N_s} \lvert \phi_l({\bf x}) \lvert^2\\
	\lvert \rho({\bf x}) \lvert=\sqrt{\frac{1}{N_s} \sum_{l=1}^{N_s} \lvert m_l({\bf x}) \lvert^2}
	\end{split}
\end{equation}

\vspace{2mm}
{ \bf Smoothness regularized reconstruction of multi-channel MS-DWI data}
\vspace{2mm}\\
The formulation in the above section works well when the data is fully sampled and not noisy. However, the recovery may be ill-conditioned when used with undersampling or when the measurements are noisy. To improve the conditioning of the reconstruction, additional smoothness penalties such as total-variation (TV) penalty can be imposed on the reconstructed images. We instead rely on an annihilation formulation for smoothness regularization, introduced in \cite{Ongie2015a,Ongie2015,Jin2015,Lee2016,Jin2015a}. These methods assume that the partial derivatives of the image vanishes on the zero-crossings of a filter $\mu$ bandlimited to $p\times p; p<r$, which translates to the space domain relation
\begin{equation}
\label{space}
\boldsymbol \nabla m(\bf x) ~\cdot~ \mu(\bf x) = \bf 0
\end{equation}
in the distributional sense. Here $\boldsymbol \nabla m(\bf x) = \begin{bmatrix}\partial_1 m({\bf x})  \hspace{3mm} \partial_2 m({\bf x}) \end{bmatrix}$ is the gradient operator and $\partial_1$ and $\partial_2$ are the partial derivative operators in x- and y- dimensions. The space domain operation \eqref{space} translates to the Fourier domain annihilation relations
\begin{eqnarray}
\widehat{\partial_1 m}[\bf k] \ast \widehat{\mu}[\bf k] &=& 0\\
\widehat{\partial_2 m}[\bf k] \ast \widehat{\mu}[\bf k] &=& 0;
\end{eqnarray}
Using the derivative property of Fourier transforms, we have $\widehat{\partial_1 m}[\mathbf k] =  j2\pi k_x~\widehat{m}[\bf k]$ and $\widehat{\partial_y m}[\mathbf k] = j2\pi k_y~\widehat{m}[\bf k]$, where $k_x$ and $k_y$ are the spatial frequencies. The above convolutions can also be replaced by multiplications using block-Hankel matrices as described in the previous section, and the resulting annihilation relations can be compactly represented in the matrix form as 
\begin{equation}
\label{kspaceannih}
\underbrace{\begin{bmatrix}
\mathbf H(\widehat{\partial_1 m})\\
\mathbf H(\widehat{\partial_2 m})
\end{bmatrix}}_{\bf G(\widehat{m})} \widehat{\boldsymbol \mu} = \mathbf 0
\end{equation}

Here, $\mathbf H(\widehat{\partial_1 m})$ and $\mathbf H(\widehat{\partial_2 m})$ are defined the same way as in the previous section and has the same dimensions. Note that the support to $\widehat\mu$ is $p\times p$, where $p<r$; one can consider $(r-p)^2$ shifts of $\widehat \mu$ that are support limited in the $r\times r$ window, all of which will satisfy \eqref{kspaceannih}. This implies that $\bf G(\widehat{m})$ is low-rank. This property was exploited to recover the signal from undersampled measurements in \cite{Ongie2015a,Ongie2015,Jin2015,Lee2016,Jin2015a}. We propose to combine the matrix liftings specified by \eqref{equation 3} and \eqref{kspaceannih} to obtain a new structured matrix:
\begin{equation}
\label{jointlifting}
\mathbf F(\bf\widehat{m}) = \begin{bmatrix} {\bf{H}}(\widehat{\partial_1 m_1}) & {\bf{H}}(\widehat{\partial_1 m_2}) & ... & {\bf H}(\widehat{\partial_1 m_{N_s}}) \\
{\bf{H}}(\widehat{\partial_2 m_1}) & {\bf{H}}(\widehat{\partial_2 m_2}) & ... & {\bf H}(\widehat{\partial_2 m_{N_s}})\end{bmatrix}
\end{equation}

See figure \ref{fig:fig3} for an illustration of the new lifted matrix $\mathbf F$, which is highly low-rank. We propose to recover the motion-compensated multishot data using the consolidated nuclear-norm minimization problem that incorporates smoothness regularization (SR):
\begin{equation}
\label{with_TV}
\hat{\tilde {\bf m}}=  \text{argmin}_{{\bf \hat{m}}} {||\bf{{\cal{A}}(\hat{ m})-\hat y}}||^2_{\ell_2} + \lambda||{\bf{\bf{F}}}(\hat{ m})||_*.
\end{equation}
The above minimization problem, which we will refer to as SR-MUSSELS, can also be solved in the same framework as that of the unconstrained formulation in Eq. [\ref{use Hankel}].

\vspace{2mm}
{ \bf Augmented Lagrangian optimization algorithm}
\vspace{2mm}

We propose to solve \eqref{use Hankel} and \eqref{with_TV} using an augmented Lagrangian (AL) iterative algorthm \cite{Bertsekas1976,Ramani2011} using variable splitting. The unconstrained minimization problems in  \eqref{use Hankel} and \eqref{with_TV} are converted to constrained forms through the introduction of the auxiliary variable $\bf D $ to get the new cost $C_1$:
\begin{equation}
\label{AL1}
C_1= {||\bf{{\cal{A}}\hat{ m}-\hat y}}||^2_{\ell_2} + \lambda||\bf D||_*  \text{ such that } {\bf{\bf{F}}}(\hat{\bf m})=\bf D.
\end{equation}
The constrained form of Eq. (\ref{AL1}) can be solved by alternatively solving the quadratic subproblem
\begin{equation}
\label{AL2}
C_2({\bf \hat m})= {||{\bf{\cal{A}}{\bf \hat m}}-{\bf \hat y}}||^2_{\ell_2} +\frac{\lambda\beta}{2}||{\bf D}-{\bf{F}}({\bf \hat m})||^2_{\ell 2} +\frac{\lambda}{2}\hat{\gamma}^T({\bf D}-{\bf F}({\bf \hat m})),
\end{equation}
and a singular value shrinkage subproblem (see Table 1 for more details).\\

{ \Large Methods}\\
{ \bf Experiments and datasets for validation}\\

We used in-vivo data collected on a 7 Tesla MR scanner using a multi-shot diffusion acquisition to test the proposed reconstruction. Several sets of data were collected on two healthy adult volunteers for the validation of the methods. In-vivo brain data were obtained in accordance with the Institutional Review Board of the University of Iowa. The volunteer was imaged on a GE 7T 950 MR scanner (GE Healthcare Waukesha, WI) with maximum gradient amplitude of 50 mT/m and slew rate of 200 T/m/sec. A 32-channel Tx/Rx coil with transmit in quadrature mode was used for imaging. Stajeskal-Tanner diffusion sequence coupled with single-shot, 2-shot, 4-shot and 8-shot readout were used for acquiring the data. DW imaging scan parameters included: b-value$=$800s/mm$^2$ and 1000s/mm$^2$;  FOV = 220mmx220mm; matrix size=128x128; slice thickness=1.7mm; and TE=57-142ms depending on b-value and number of shots. All data consisted of six diffusion weighted acquisitions and one non-diffusion weighted acquisition. To simulate under-sampled multi-shot acquisition, the 4-shot data was retrospectively under-sampled by a factor of 2.\\

Figure \ref{fig:fig6} demonstrates the effect of TE on image quality of the DWIs collected at 7 Tesla. Because of the short T2 relaxation times at higher field strengths, shorter TEs becomes extremely desirable for studies at higher field strengths. 4-shot or higher number of shots provide good SNR and provide a significant reduction in the TE. As a result, susceptibility related artifacts are also minimized. The MUSE method has been shown to be effective in unaliasing the motion-induced ghosting artifacts for a 4-shot DWI acquisition when good phase estimates are available \cite{Chen2013}. For this reason, we implemented the MUSE method and we compare the proposed reconstruction with this method. For the purpose of comparing with the proposed regularized reconstruction, which included a smoothness constraint, the MUSE algorithm also included an iterative reconstruction using the TV constraint.  \\

Coil sensitivity information was required for the reconstruction using the proposed method as well as the other methods compared in this work. The coil sensitivity maps were estimated from the non-diffusion weighted images by combining the k-space data from all the shots into a single data matrix and performing an inverse Fourier transform and using a sum-of-squares (SOS) scheme \cite{Chen2013}. Note that any Nyquist ghost artifacts resulting from odd-even shifts of the EPI acquisition needed to be corrected prior to this step. Residual ghosting present in the sensitivity maps resulted in residual artifacts in the final reconstruction. The data used in this manuscript were corrected for odd-even EPI shifts before computing the coil sensitivities using standard reference-scan based methods, which were not fully effective for multi-shot acquisitions. Thus, some residual ghosting was still visible in the images. As noted previously, in addition to the coil sensitivity maps, the MUSE algorithm required the motion-induced phase maps corresponding to each shot to reconstruct the DWI. Using coil sensitivity maps already computed, an estimate of the motion-induced phase maps were obtained by using a TV-regularized reconstruction for each of the k-space shots as proposed in Chen et al \cite{Chen2013}\\

\vspace{-2mm}
{ \bf Multi-shot DWI reconstruction without explicit phase estimation}
 \vspace{2mm}\\
In the first experiment, we show the capability of MUSSELS in recovering the unaliased k-space data from a 4-shot acquisition. The measured 4-shot k-space data were channel combined and stacked into a 3D matrix as shown in figure 2 and the Hankel matrix was computed. An 8 x 8 window was used for the Hankel matrix construction. 
In the second experiment, we demonstrate the utility of the smoothness-regularized version of MUSSELS (SR-MUSSELS). We show that SR-MUSSELS can be used to reconstruct images from ill-conditioned data reconstruction problems such as an 8-shot MS-DW data or under-sampled MS-DW data. A window size of 12 x 12 was used in this case for the Hankel matrix construction and the low-rank property was imposed on the taller Hankel matrix shown in figure (\ref{fig:fig3}).\\

We used the AL scheme for the recovery of the data, the pseudo-code for which is provided in Table 1. The proposed implementation is fast, and the speed is determined by the filter size. For example, a 128x128 matrix size 4-shot 32-channel data with an 8x8 filter took 25 secs to reconstruct the unaliased image on an Intel i7-4770, 3.4GHz CPU with 8GB RAM using Matlab. The maximum filter size that we used in our experiments was 12 x 12; however, an 8 x 8 filter gave comparable results in all the cases. \\

{ \Large Results}\\

\vspace{-2mm}
{ \bf Unaliasing without phase estimates}
\vspace{2mm}\\
Figure \ref{fig:mussel} shows the reconstruction results using the proposed scheme MUSSELS from a 4-shot acquisition. For comparison, the results from a conventional SENSE reconstruction is also included. Both of these methods use only the coil sensitivity information and do not use motion-induced phase maps. As evident from the figure, conventional SENSE reconstruction exhibits the motion-induced aliasing artifacts while the MUSSELS algorithm successfully removes the motion-induced artifacts from the DWIs.\\

\vspace{-2mm}
{ \bf Comparison to methods that use phase estimates}
\vspace{2mm}\\
Next we compare the performance of MUSSELS with the standard method that use motion-induced phase estimates to achieve unaliasing of the DWIs. Figure \ref{fig:compare4shot_muse_mussels} shows two sample DWIs corresponding to two different diffusion directions reconstructed from the 4-shot acquisition using a regular SENSE method, the MUSE method and MUSSELS. As can be seen from the figure, MUSE is successful in unaliasing the first DWI (Fig. \ref{fig:compare4shot_muse_mussels} DWI $\#1$), but not the the second DWI (Fig. \ref{fig:compare4shot_muse_mussels} DWI $\#2$). The phase maps used in the recovery of the DWIs for MUSE reconstruction are also included in the figure. A six direction diffusion scheme was used here to highlight the errors originating from the reconstruction in the computation of the fiber directions. The color coded FA maps given in the last row shows the smearing of fiber orientation in the case of MUSE whereas the fiber orientations are accurately recovered using MUSSELS. Note that the same coil sensitivity maps were used in all the reconstructions. \\

\vspace{-2mm}
{ \bf Regularized reconstruction of  8-shot data}
\vspace{2mm}\\
The recovery of 8-shot diffusion data can be challenging due to the fewer number of sampled data points present in each shot.  Because of the absence of a fully sampled center k-space, the phase estimates obtained from the shots can be noisy. Hence the reconstructions that depend on the phase estimates may not achieve good unaliasing in such situations. MUSSELS with no additional constraint will generate noisy results since it is hard to characterize the k-space filter corresponding to the phases from the smaller number of samples present in the compact filter. However the reconstruction can be stabilized by adding extra constraints. By using the smoothness regularized reconstruction provided in Eq [\ref{with_TV}], we show that we can significantly improve the unaliasing for under-sampled acquisitions as well as for data collected with a high number of shots. Figure \ref{fig:compare8shot_muse_mussels} shows the reconstruction of the 8-shot MS-DWI data using unregularized and regularized methods. To introduce a similar smoothness regularizer into the MUSE framework, a total-variation based regularization was added to the MUSE reconstruction as well. The estimates of phase used in MUSE reconstruction for one of the images (indicated by the colored boxes) are also provided in the figure. The top row shows the reconstruction using MUSE with no TV during the image reconstruction. The left figure used low regularization for phase estimation (the regularization parameter was chosen to give the best reconstruction for the boxed image in  \ref{fig:compare8shot_muse_mussels}(a)) while the right figure used higher regularization for phase estimation. The corresponding phase images are given in the panel on right; it shows that the quality of the phase images control the quality of the DWI reconstruction. The second row shows the MUSE reconstruction with TV built into the iterative MUSE implementation. However, as can be appreciated from the images, the TV reconstruction also gets only as better as dictated by the noise in the phase since these reconstructions are inherently phase-based. The third row shows the MUSSELS reconstruction without TV (left) and with TV (right). As expected the images are noisy for the unregularized case. However using the smoothness regularized MUSSELS, we can recover the MS-DWI data reasonably well.\\

{ \bf Regularized reconstruction for under-sampled multi-shot DWI}\\
The previous experiments demonstrate the utility of MUSSELS to recover the fully sampled MS-DWI data. Here we show here that the regularized version of MUSSELS can be used to recover under-sampled MS-DWI data as well. For this purpose the 4-shot MS-DWI data was first under-sampled uniformly by skipping every other k-space lines from each of the shots. The top two rows of figure \ref{fig:compareus_muse_mussels} shows the reconstruction of the 4-shot under-sampled MS-DWI data using TV-MUSE and SR-MUSSELS. This experiment highlight the situation where phase-based methods will fail. The inadequate unalising is evident in all DWIs reconstructed by MUSE. The regularized MUSSELS has performed reasonably well with significantly fewer artifacts seen visually in the reconstructed images. The color-coded FA maps generated from the six DWIs highlights the errors in the individual reconstructions. Interestingly, if the under-sampling pattern is changed slightly, the performance of both the methods improve significantly. The bottom rows of figure \ref{fig:compareus_muse_mussels} shows the results of reconstruction where a non-uniform under-sampling pattern was employed. Specifically, the  center k-space lines of each of the shots were kept intact. The improvement in the reconstruction results can be appreciated from the DWIs as well as the color coded FA maps. This behavior is not surprising and adds evidence to the fact that reconstructions from a slightly non-uniform under-sampling patterns provides more reliable results than a strictly uniform under-sampling pattern while using sparsity/low-rank -based reconstructions \cite{Haldar2015}. \\

{\Large Discussion}\\  

 Recently structured low rank matrix completion problem had been extensively studied in the context of calibration-less parallel imaging reconstruction. Coupled with the theories from finite rate of innovations, later works showed that annihilation relations in the frequency domain can be exploited to learn shift-invariant filters that can recover any missing k-space data in a local neighborhood. Recently, it was also shown that such annihilating filters provide a flexible and generalized framework for reconstruction of MR images and can easily incorporate other constraints such as sparsity into the formulation.\\
In this work, we modeled the smooth slowly varying phase of the MS-DWI data by a shift invariant k-space filter. By deriving an annihilation formulation based on theses filters, we derived a reconstruction framework based on structured low-rank recovery that could learn the filter implicitly to recover the missing k-space data. This lead us to a new formulation that can recover diffusion weighted images from a multi-shot acquisition with inherent motion-compensation.  \\
While the method introduced here shares its underlying principles with other calibration-less methods such as SAKE and LORAKS, it should pointed out here that these methods cannot be directly extended to the recovery of multi-shot diffusion data by a low-rank minimization along the shot-dimension. The fundamental difference here is the knowledge and utilization of the coil sensitivity information. Since SAKE and LORAKS were designed to recover missing k-space data resulting from the use of parallel imaging calibration-free, the coil sensitivity information is absent in its formulation. Consequently, not all patterns of missing k-space data could be recovered using those methods, especially the uniformly unacquired k-space points such as that we have in a multi-shot diffusion imaging scheme. Since a non-diffusion weighted image is inevitably acquired in these imaging sessions, we do have access to the coil sensitivity information. This gives us the capability to recover the missing uniform k-space samples in a multi-shot diffusion acquisition. Note that since we recover the k-space data for each shot, we can recover the final diffusion weighted image without having to know the motion-induced phase that varies between shots. Thus, the formulation that we introduced here, which we call MUSSELS,  is also a "calibration-free" method since we do not require the motion-induced phase anymore to recover the final diffusion weighted images. \\
As shown from our results above, the proposed method can enable direct motion-compensated reconstruction for short-readout EPI acquisitions such as multi-shot acquisitions which in turn can be used more readily for routine imaging. The proposed method can greatly benefit the acquisition of high resolution DWI as well as high quality DWI at ultra-high field strengths. This method is easily adopted to non-Cartesain acquisitions as well. \\

In conclusion, we proposed a fast and robust reconstruction scheme for fully sampled and under-sampled multi-shot diffusion data recovery that does not rely on motion induced phase estimates or navigator data. 

\vspace{2mm}
{\bf{{Acknowledgements}}}
\vspace{2mm}

Financial support for this study was provided by NIH grants 1R01EB019961-01A1 and NIH T32 MH019113-23.

\newpage
{\bf{Appendix: }}\\
We can derive a different set of annihilating filters based on the coil sensitivities similar to that we derived above using the phase. Writing the DWI for shot $l$ as the multiplication of the object $\rho(\bf{x})$ with the coil sensitivities $s({\bf{x}})$, we have $m_{il}({\bf{x}})=\rho({\bf{x}})s_i({\bf{x}})$; $i=1:N_c$. Thus for coils $i,j$ of shot $l$, we have:
	\begin{equation}
	\label{imp}
	{m}_{il}({\bf{x}})={\rho}({\bf{x}})s_i({\bf{x}})
	\end{equation}
	 \begin{equation}
	 \label{imp2}
	 {m}_{jl}({\bf{x}})={\rho}({\bf{x}})s_j({\bf{x}}).
	 \end{equation}
Multiplying [\ref{imp}] by $s_j({\bf{x}})$ and [\ref{imp2}] by  $s_i({\bf{x}})$ and subtracting we get the following: 	\begin{equation}
	{m}_{il}({\bf{x}}) s_j({\bf{x}})-{m}_{jl}({\bf{x}}) s_i({\bf{x}})=0 \text{\hspace{5mm}}      \forall i,j.		 \end{equation}
Equivalently, in Fourier domain \cite{Morrison2007},
	\begin{equation}
	\label{eq_7}
	\widehat{m_{il}}[{\bf{k}}]*\widehat s_j[{\bf{k}}]-\widehat{m_{jl}}[{\bf{k}}]*\widehat s_i[{\bf{k}}]=0 \text{\hspace{5mm}}      		\forall i,j.
	\end{equation}
Replacing the convolution operations in Eq [\ref{eq_7}] as multiplication using Toeplitz/Hankel matrices, we get \begin{equation}
	\label{eq_8}
	{\bf{H}}(\widehat{m_{il}}) \cdot \widehat s_j [{\bf{k}}]- {\bf{H}}(\widehat{m_{jl}}) \cdot \widehat s_i[{\bf{k}}]=0.
	\end{equation}
Since this is true for all shots $l=1:N_s$, we can derive a set of conditions which can be written in matrix form as:
	\begin{equation}
	\label{Sen_for}
	\underbrace{\begin{bmatrix} \widehat s_2 & -\widehat s_1& 0&... &0 \\0&\widehat s_3& -\widehat s_2&... &0\\ \vdots\\ \widehat s_3 & 0 &  -\widehat s_1 &... & 0 \\ \vdots \\0 & 0 &... & \widehat s_{N_c} &  -\widehat s_{N_c-1}\end{bmatrix}}_{{\bf{\widehat S({\bf{k}})}}}\underbrace{\begin{bmatrix} {\bf{H}}(\widehat{m}_{11}) & {\bf{H}}(\widehat{m}_{12}) ... & {\bf{H}}(\widehat{m}_{1N_s}) \\ \\{\bf{H}}(\widehat{m}_{21}) & {\bf{H}}(\widehat{m}_{22}) ... & {\bf{H}}(\widehat{m}_{2N_s}) \\ \vdots \\{\bf{H}}(\widehat{m}_{N_c1}) & {\bf{H}}(\widehat{m}_{N_c2}) ... & {\bf{H}}(\widehat{m}_{N_cN_s}) \end{bmatrix}}_{{\bf{H}}_2(\widehat{\bf m})} =0 \Rightarrow  {\bf{\widehat S}}{\bf{H}}_2(\widehat{\bf m}) = 0.
	 \end{equation}
Since we have access to the non-diffusion weighted image for all imaged slices, which are not affected by the shot-to-shot phase variations, we can compute the coil sensitivity images $s({\bf{x}})$ using a sum-of-squares (SOS) method or the k-space filters $\widehat s[{\bf{k}}]$ with the center k-space lines of the non-diffusion weighted data acting as the ACS lines using methods such as ESPIRiT \cite{Uecker2014}. The matrix ${{\bf{\widehat S({\bf{k}})}}}$ as defined in [\ref{Sen_for}] can have 
$\comb{N_c}{2}$
 rows which implies that the left null-space of the block-Hankel matrix ${{\bf{H}}_2(\widehat{\bf m})}$ has high dimensionality. 
With the above filter formulation, we can pose the recovery of the multi-shot k-space data $\tilde {\bf m}$ as the constrained reconstruction problem:  
	\begin{equation}
	\label{compli_recon0}
	\begin{split}
\widehat{\tilde {\bf m}}=  \mathrm{min} \hspace{2mm} rank(\bf{\bf{H_1}}({\bf{\widehat{m}}})) 
\text{   subject to   }\\ \bf{\widehat{S}}{\bf{H}}_2({\bf{\widehat{m}}}) = 0 , \text{ and }\\ {\mathbb A}{\bf{\widehat{m}}}={\bf{\widehat{y}}} 
	\end{split}
	\end{equation}
	The last term imposes the data consistency to the measured multi-channel multi-shot k-space data, $\bf{\hat y}$, of dimension $N_1\times N_2\times N_c\times N_{s}$. The operator $\bf{{\mathbb{A}}}$ applies the mask corresponding to the different multi-shot sampling locations. The equivalent unconstrained reconstruction problem can be written as
	\begin{equation}
	\label{compli_recon}
\widehat{\tilde {\bf m}}=  \argmin_{{\bf{\widehat{m}}}} {||{\mathbb A}{\bf{\widehat{m}}}-{\bf{\widehat{y}}}||^2_{\ell_2}} + \lambda_1||\bf{\bf{H_1}}({\bf{\widehat{m}}})||_* + \lambda_2 ||{\bf{\widehat{S}}{\bf{H}}}_2({\bf{\widehat{m}}})||^2_{\ell_2} ,
	\end{equation}
where $ \lambda_1$ and  $\lambda_2$ are two regularization parameters.  In practice, numerically solving the above reconstruction problem is highly computation-intensive mainly because of the size of the Hankel matrix ${\bf{\bf{H}}_2(\widehat{m})}$, especially with high number of coils. Hence the SENSE-based method is  adopted in this work.

\newpage
{\bf{Legends: }}\\
Fig 1: (a) A 4-shot acquisition illustrated. (b) The k-space data matrix of the 4-shot DWI acquisition. The solid circles and the hollow circles represent the acquired and unacquired k-space samples during each shot respectively.\\

Fig 2: Illustration of the matrix lifting:  ${\bf{\hat{{m}}}}$ is the k-space data matrix of a given DWI comprising of data from  the different shots of the DWI. A sliding window of size $r \times r$ as marked by the red dotted box generates the rows of the block-Hankel matrix  $\bf{\bf{H}}({\bf{\hat{{m}}}})$ by vectorizing the elements in the red block.\\

Fig 3: Illustration of joint matrix lifting for multi-shot data: The Fourier coefficients of the partial derivatives along the x-dimension and y-dimensions are obtained by multiplication using $-j2\pi k_x$ and $-j2\pi k_x$, respectively. The block Hankel matrices of the each partial derivative are generated and and stacked as shown.\\

Fig 4: Effect of long echo times at 7T demonstrated on non-diffusion weighted images collected using different number of shots for a 128 x 128 acquisition matrix. Due to the shortened T2 relaxation times, high signal drop-out are observed in many regions of the single and two-shot acquisitions as pointed out by the circles. The SNR computed from the ROIs as a function of the number of shots are shown in the last column. No parallel imaging acceleration was employed in these acquisitions. However, with single-shot imaging, it is common to employ parallel imaging acceleration of at least 2, in which case, the TE becomes comparable to the 2-shot case in column two.\\

Fig 5: DWIs reconstructed from a 6-direction (b=1000$s/mm^2$) 4-shot acquisition using conventional SENSE method (a) and the proposed MUSSELS method (b). Both methods use the coil sensitivity information only and do not use motion-induced phase estimates in the reconstruction. Conventional SENSE cannot achieve motion unaliasing while MUSSELS can recover the unaliased DWIs.\\

Fig 6: DWI $\#1$ and DWI $\#2$ are two diffusion weighted images corresponding to two different diffusion directions. The first row shows (a) conventional SENSE,  (b) MUSE and (c) MUSSELS reconstruction for DWI $\#1$.  While MUSE and MUSSELS effectively unaliased the ghosting artifacts for DWI $\#1$, SENSE reconstruction shows residual artifacts. The second row shows the motion-induced phase maps that were estimated as part of MUSE reconstruction for the four shots of the DWI shown in the first row. The third row shows (d) SENSE,  (e) MUSE and (f) MUSSELS reconstruction for DWI $\#2$. In this case, the MUSE reconstruction was not effective in unaliasing the ghosting artifacts.  SENSE reconstruction also shows the ghosting whereas MUSSELS shows good unaliasing. The fourth row shows the phase maps used by MUSE for the DWI shown in the third row. The final row shows the color-coded fractional anisotropy maps recovered using SENSE (g), MUSE (h) and MUSSELS (i). A six direction DTI acquisition was used to highlight the errors in tensor estimation due to the residual aliasing present in the six DWIs.\\

Fig 7: 8 shot fully sampled data. The six DWIs reconstructed using various methods are shown. (a) and (f) shows MUSE reconstruction with no TV during the image reconstruction. They use TV in the estimation of estimate phase. (f) used a higher regularizer than (a) during the estimation of phase. As a result, the DWIs in (f) are denoised while the unaliasing becomes imperfect. (b) and (g) shows MUSE reconstruction with TV regularization added to the image reconstruction step also with (b) using the same phase as (a) and (g) using the same phase as (f). Inclusion of TV to the image reconstruction step has improved the MUSE reconstruction slightly however, the results are dictated by the noise in the phase estimates. (k) and (m) shows the results of MUSSELS without and with SR respectively. It can be appreciated from the color coded FA maps (e) and (l) that MUSSELS without SR achieves the better quality of reconstruction than the TV-MUSE and the results can be further improved by using SR-MUSSELS.\\

Fig 8: 4 shot under-sampled MS-DW data. (a) shows the TV-MUSE and SR-MUSSELS reconstruction of 3 DWIs reconstructed from uniformly under-sampled data. The under-sampling pattern is shown on the right side. (b) shows the unregularized as well as regularized MUSE and MUSSELS reconstruction of 3 DWIs reconstructed from non-uniformly under-sampled data. The under-sampling pattern used is included which shows that the center-k-space lines were kept intact for all the shots. The unregularized reconstructions also performed reasonably well with the non uniform under-sampling pattern.\\

Table 1: Augmented Lagrangian algorithm for solving SR-MUSSELS.\\

\newpage
{\small
\bibliographystyle{unsrtnat} 
\bibliography{ref_dti_cs_mrm}}

\newpage
\begin{figure}
\includegraphics[trim = 0mm 0mm 0mm 0mm, clip, width=.85\textwidth]{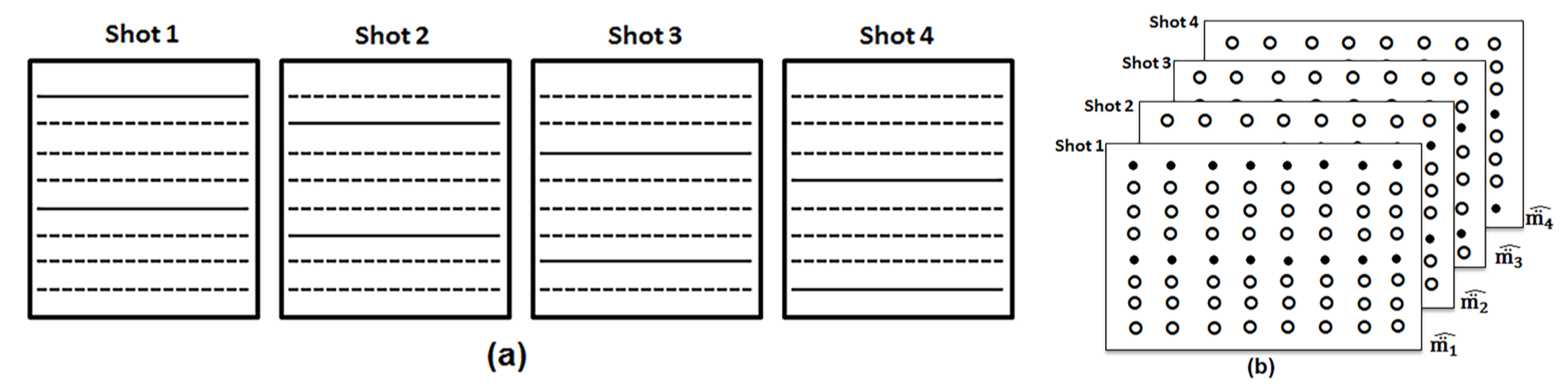}
\caption{(a) A 4-shot acquisition illustrated. (b) The k-space data matrix of the 4-shot DWI acquisition. The solid circles and the hollow circles represent the acquired and unacquired k-space samples during each shot respectively. }
\label{fig:fig1}
\end{figure}
\clearpage

\newpage
\begin{figure}[b!]
\includegraphics[trim = 0mm 0mm 0mm 0mm, clip, width=.75\textwidth]{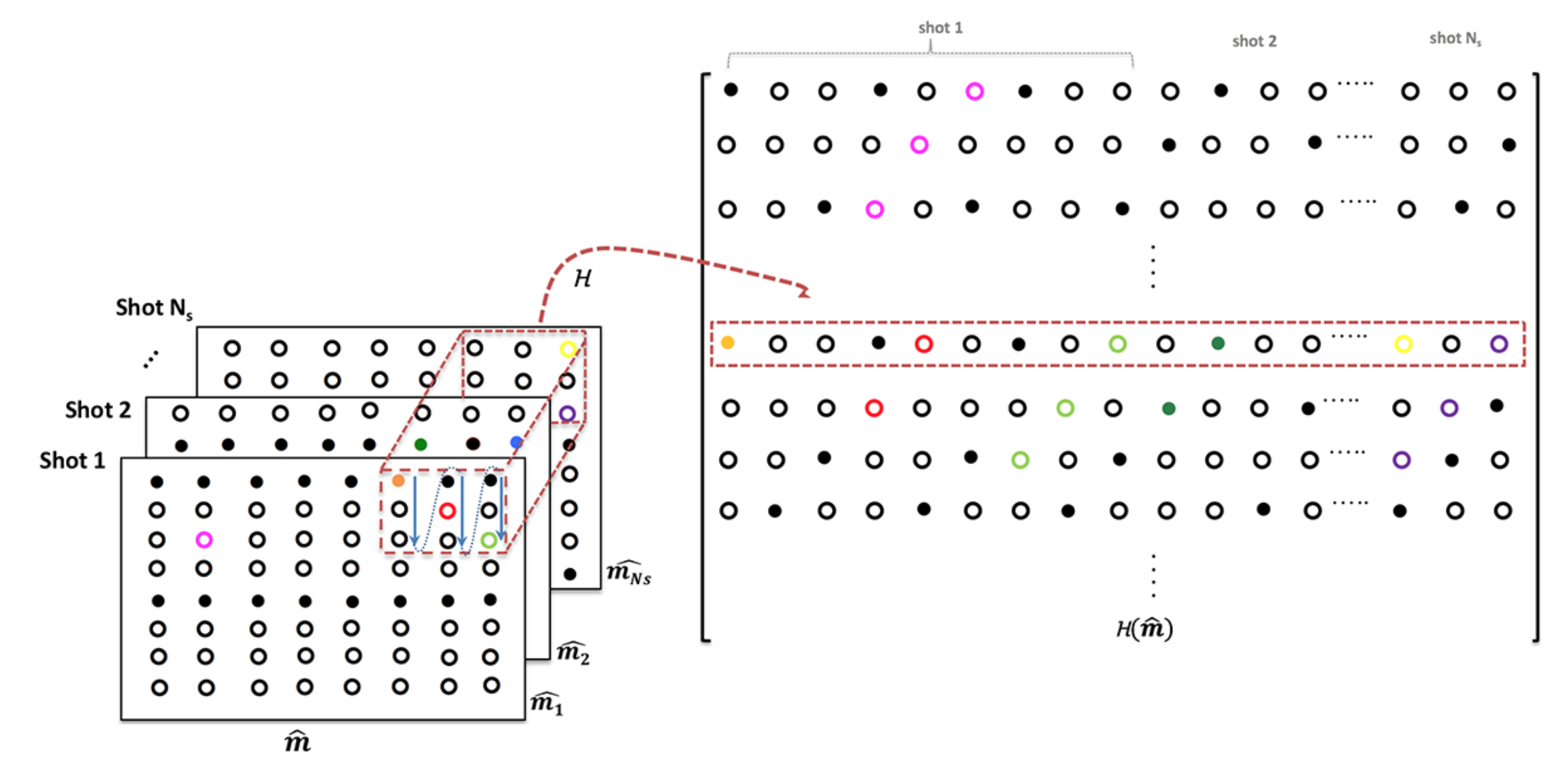}
\caption{Illustration of the matrix lifting:  ${\bf{\widehat{{m}}}}$ is the k-space data matrix of a given DWI comprising of data from  the different shots of the DWI. A sliding window of size $r \times r$ as marked by the red dotted box generates the rows of the block-Hankel matrix  $\bf{\bf{H}}({\bf{\widehat{{m}}}})$ by vectorizing the elements in the red block. }
\label{fig:fig2}
\end{figure}
\clearpage

\newpage

\begin{figure}[b!]
\includegraphics[trim = 0mm 0mm 0mm 0mm, clip, width=.75\textwidth]{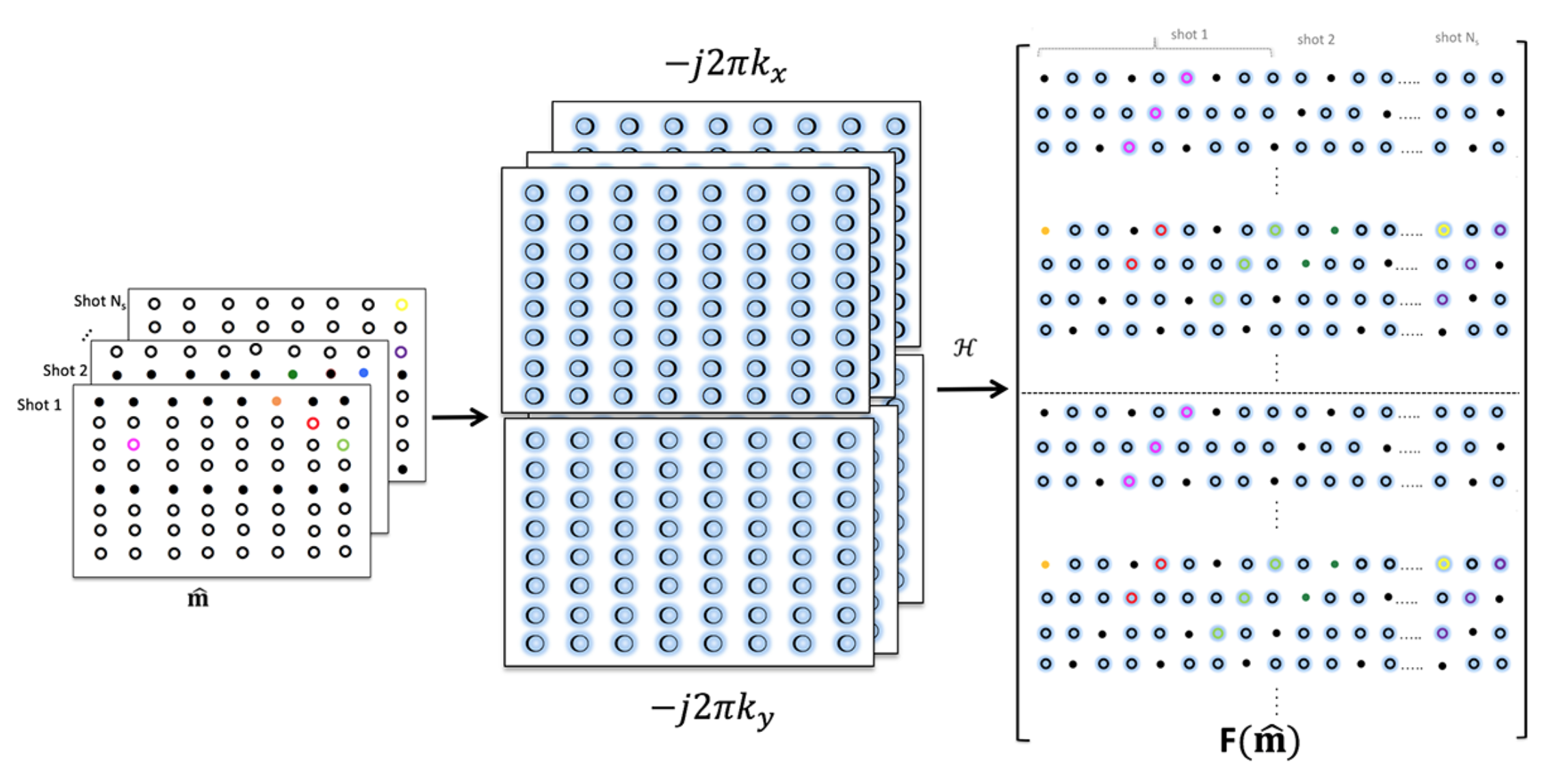}
\caption{Illustration of joint matrix lifting for multi-shot data: The Fourier coefficients of the partial derivatives along the x-dimension and y-dimensions are obtained by multiplication using $-j2\pi k_x$ and $-j2\pi k_x$, respectively. The block Hankel matrices of the each partial derivative are generated and and stacked as shown. }
\label{fig:fig3}
\end{figure}
\clearpage

\newpage

\begin{figure}
\includegraphics[trim = 0mm 0mm 0mm 0mm, clip, width=.88\textwidth]{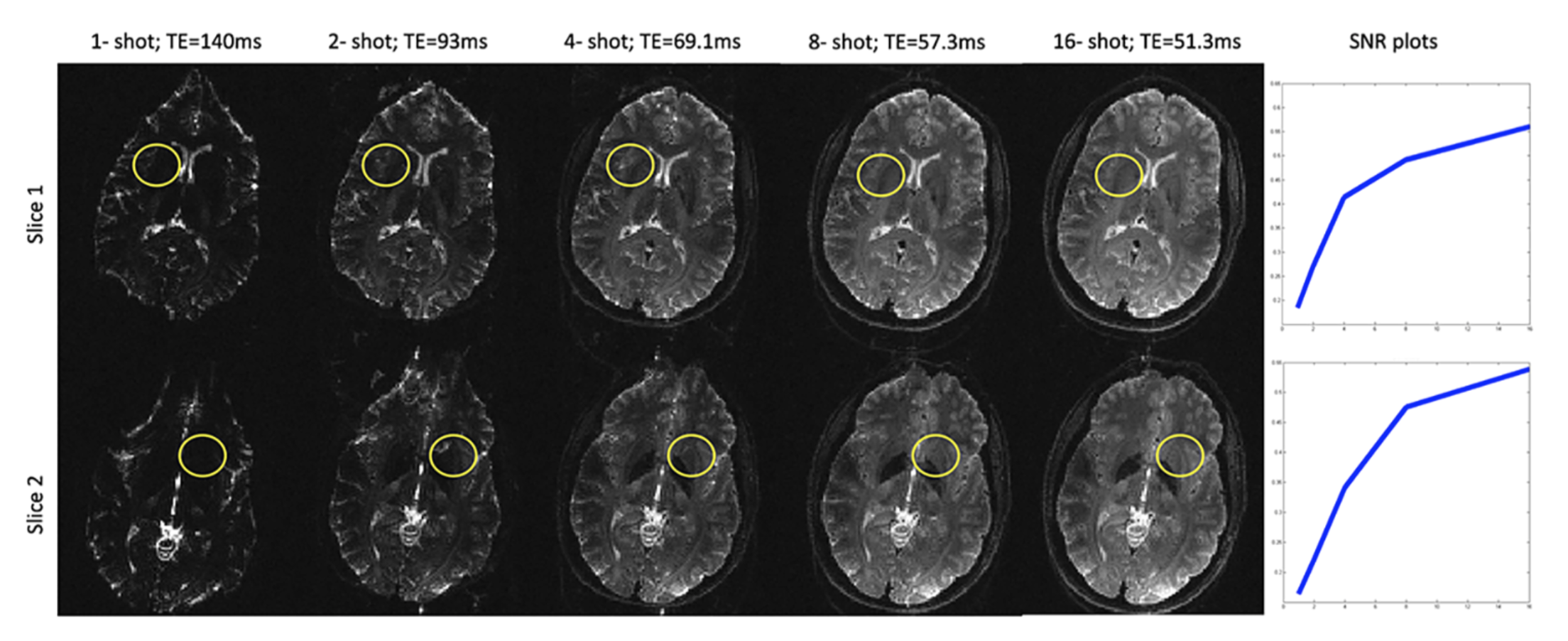}
\caption{Effect of long echo times at 7T demonstrated on non-diffusion weighted images collected using different number of shots for a 128 x 128 acquisition matrix. Due to the shortened T2 relaxation times, high signal drop-out are observed in many regions of the single and two-shot acquisitions as pointed out by the circles. The SNR computed from the ROIs as a function of the number of shots are shown in the last column. No parallel imaging acceleration was employed in these acquisitions. However, with single-shot imaging, it is common to employ parallel imaging acceleration of at least 2, in which case, the TE becomes comparable to the 2-shot case in column two.}
\label{fig:fig6}
\end{figure}
\clearpage

\newpage

\begin{figure}
\centering
\includegraphics[trim = 0mm 0mm 0mm 0mm, clip, width=0.85\textwidth]{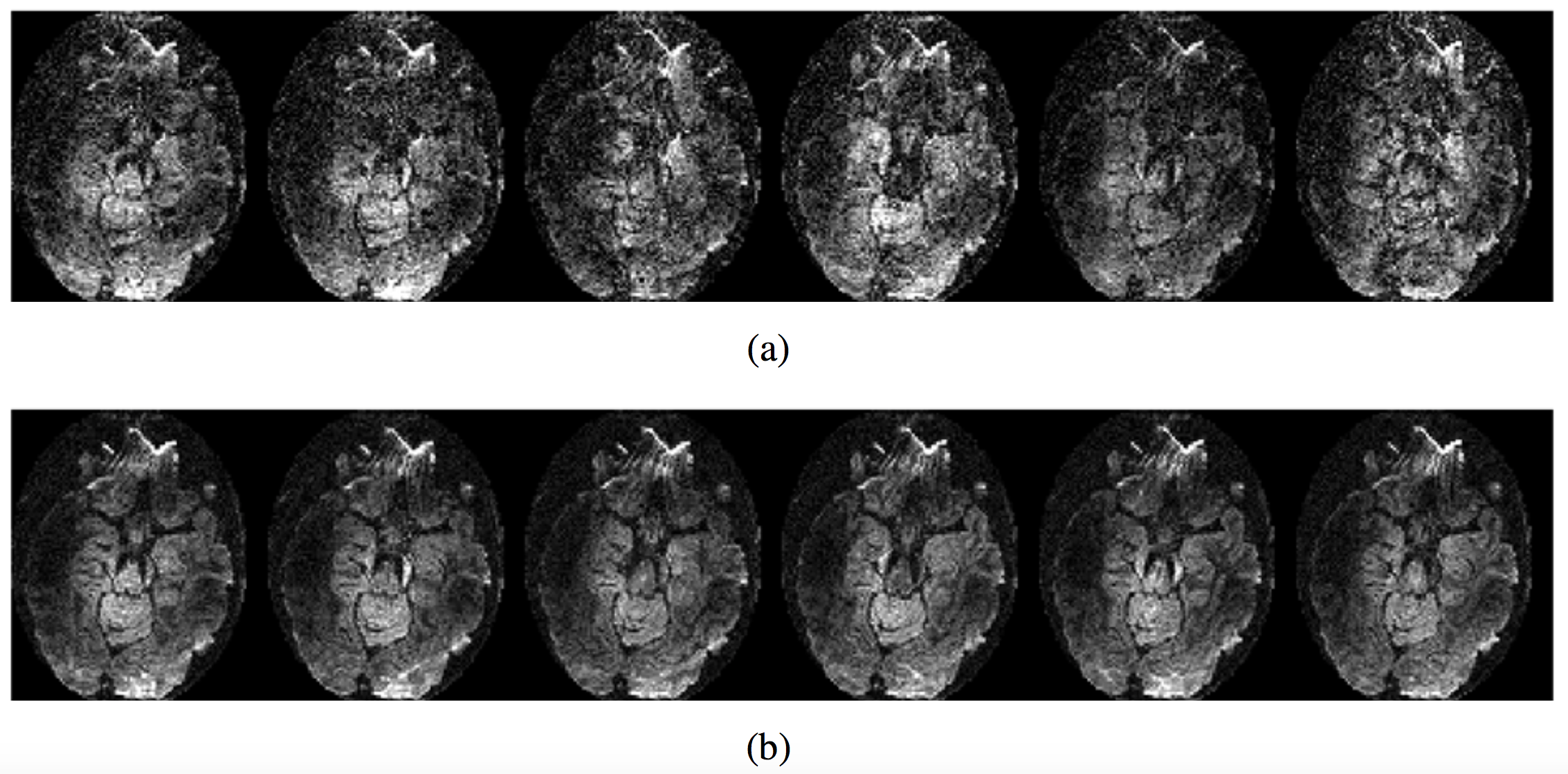}
\caption{DWIs reconstructed from a 6-direction (b=1000$s/mm^2$) 4-shot acquisition using conventional SENSE method (a) and the proposed MUSSELS method (b). Both methods use the coil sensitivity information only and do not use motion-induced phase estimates in the reconstruction. Conventional SENSE cannot achieve motion unaliasing while MUSSELS can recover the unaliased DWIs.}
\label{fig:mussel}
\end{figure}
\clearpage

\newpage

\begin{figure}
\centering
\includegraphics[trim = 0mm 0mm 0mm 0mm, clip, width=.75\textwidth]{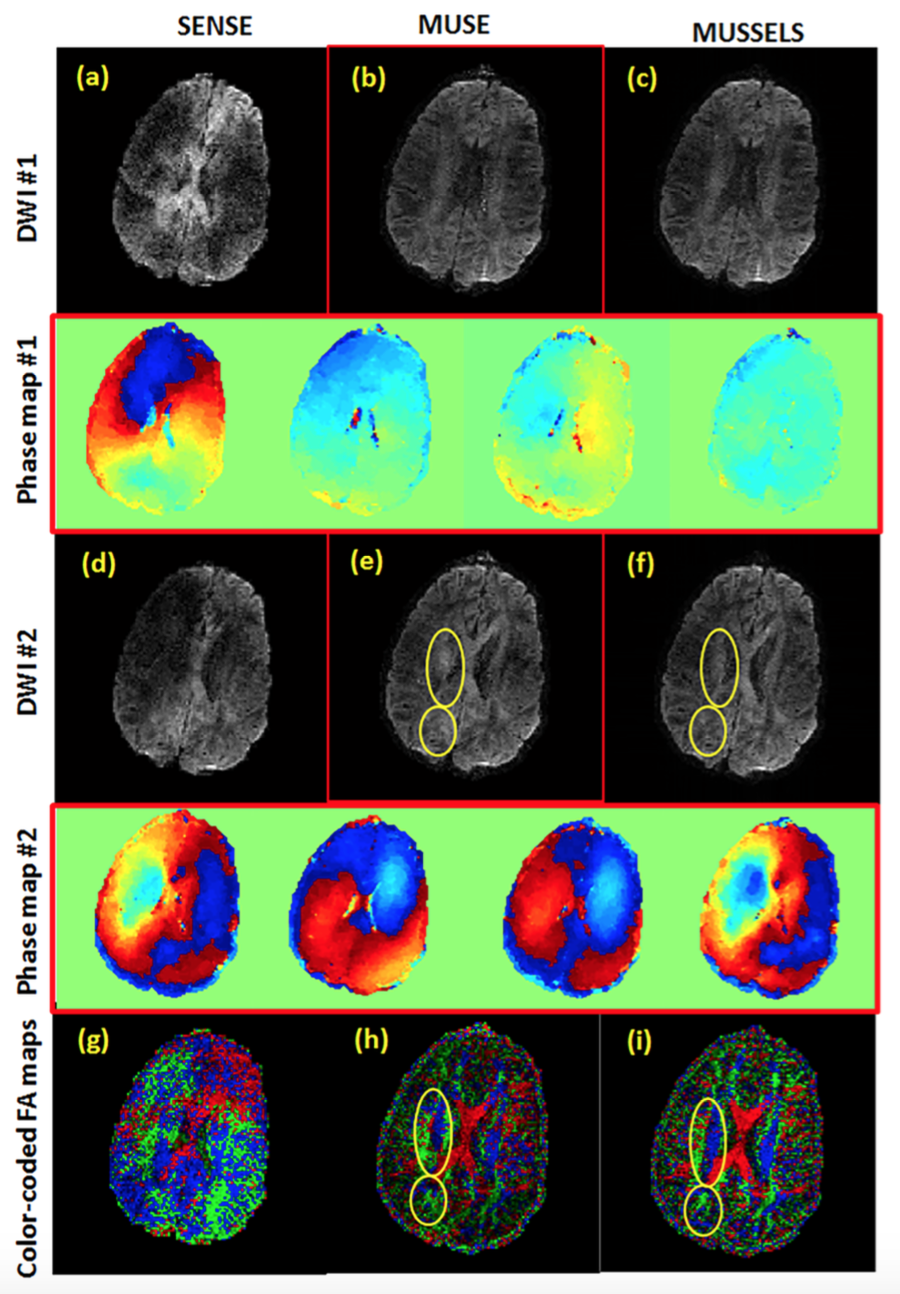}
\caption{DWI $\#1$ and DWI $\#2$ are two diffusion weighted images corresponding to two different diffusion directions. The first row shows (a) conventional SENSE,  (b) MUSE and (c) MUSSELS reconstruction for DWI $\#1$.  While MUSE and MUSSELS effectively unaliased the ghosting artifacts for DWI $\#1$, SENSE reconstruction shows residual artifacts. The second row shows the motion-induced phase maps that were estimated as part of MUSE reconstruction for the four shots of the DWI shown in the first row. The third row shows (d) SENSE,  (e) MUSE and (f) MUSSELS reconstruction for DWI $\#2$. In this case, the MUSE reconstruction was not effective in unaliasing the ghosting artifacts.  SENSE reconstruction also shows the ghosting whereas MUSSELS shows good unaliasing. The fourth row shows the phase maps used by MUSE for the DWI shown in the third row. The final row shows the color-coded fractional anisotropy maps recovered using SENSE (g), MUSE (h) and MUSSELS (i). A six direction DTI acquisition was used to highlight the errors in tensor estimation due to the residual aliasing present in the six DWIs. }
\label{fig:compare4shot_muse_mussels}
\end{figure}
\clearpage

\newpage

\begin{figure}
\centering
\includegraphics[trim = 0mm 0mm 0mm 0mm, clip, width=0.86\textwidth]{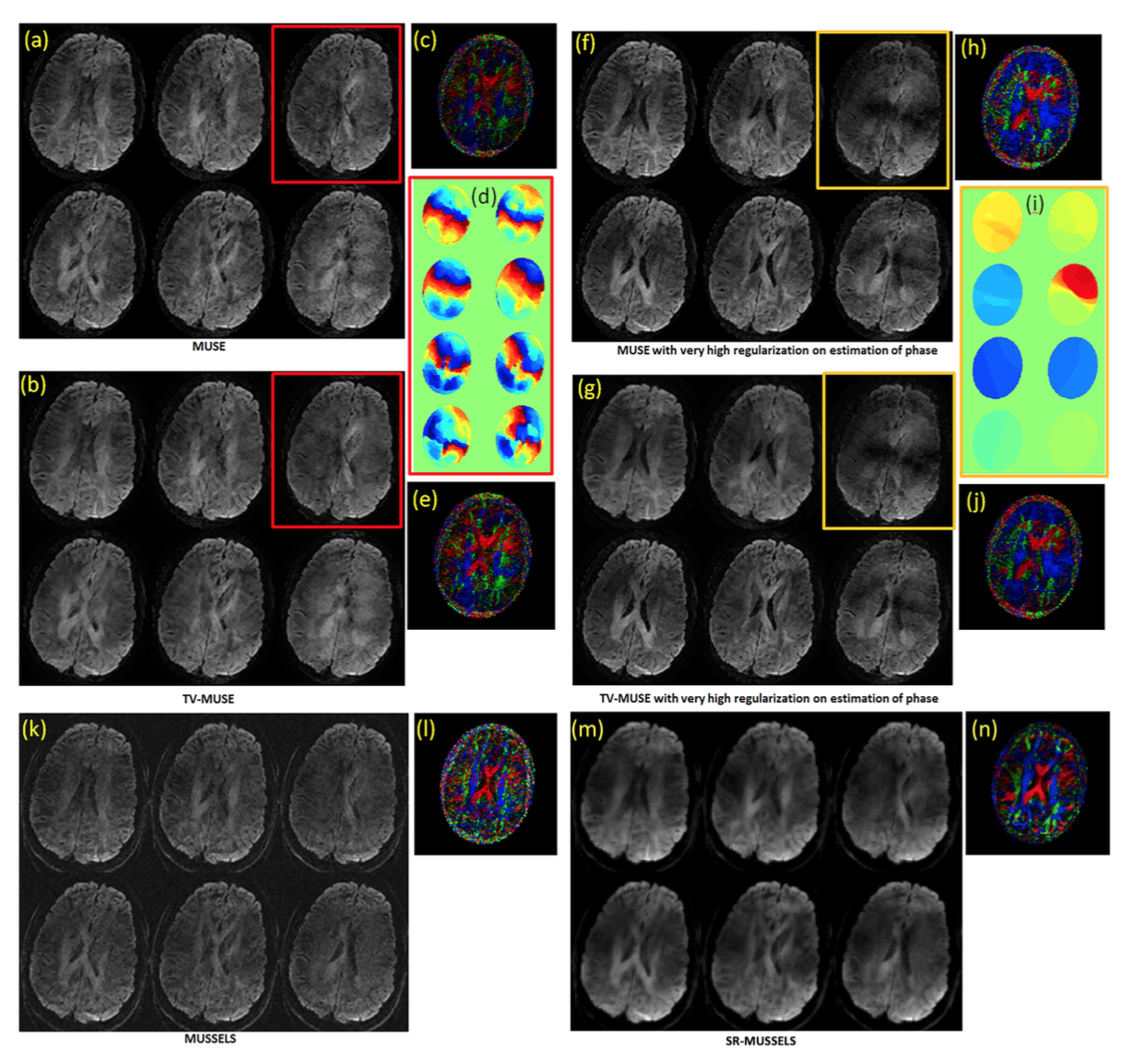}\\
\hspace{-10mm}\caption{8 shot fully sampled data. The six DWIs reconstructed using various methods are shown. (a) and (f) shows MUSE reconstruction with no TV during the image reconstruction. They use TV in the estimation of estimate phase. (f) used a higher regularizer than (a) during the estimation of phase. As a result, the DWIs in (f) are denoised while the unaliasing becomes imperfect. (b) and (g) shows MUSE reconstruction with TV regularization added to the image reconstruction step also with (b) using the same phase as (a) and (g) using the same phase as (f). Inclusion of TV to the image reconstruction step has improved the MUSE reconstruction slightly however, the results are dictated by the noise in the phase estimates. (k) and (m) shows the results of MUSSELS without and with SR respectively. It can be appreciated from the color coded FA maps (e) and (l) that MUSSELS without SR achieves the better quality of reconstruction than the TV-MUSE and the results can be further improved by using SR-MUSSELS. }
\label{fig:compare8shot_muse_mussels}
\end{figure}
\clearpage

\newpage

\begin{figure}
\centering
\includegraphics[trim = 0mm 0mm 0mm 0mm, clip, width=.84\textwidth]{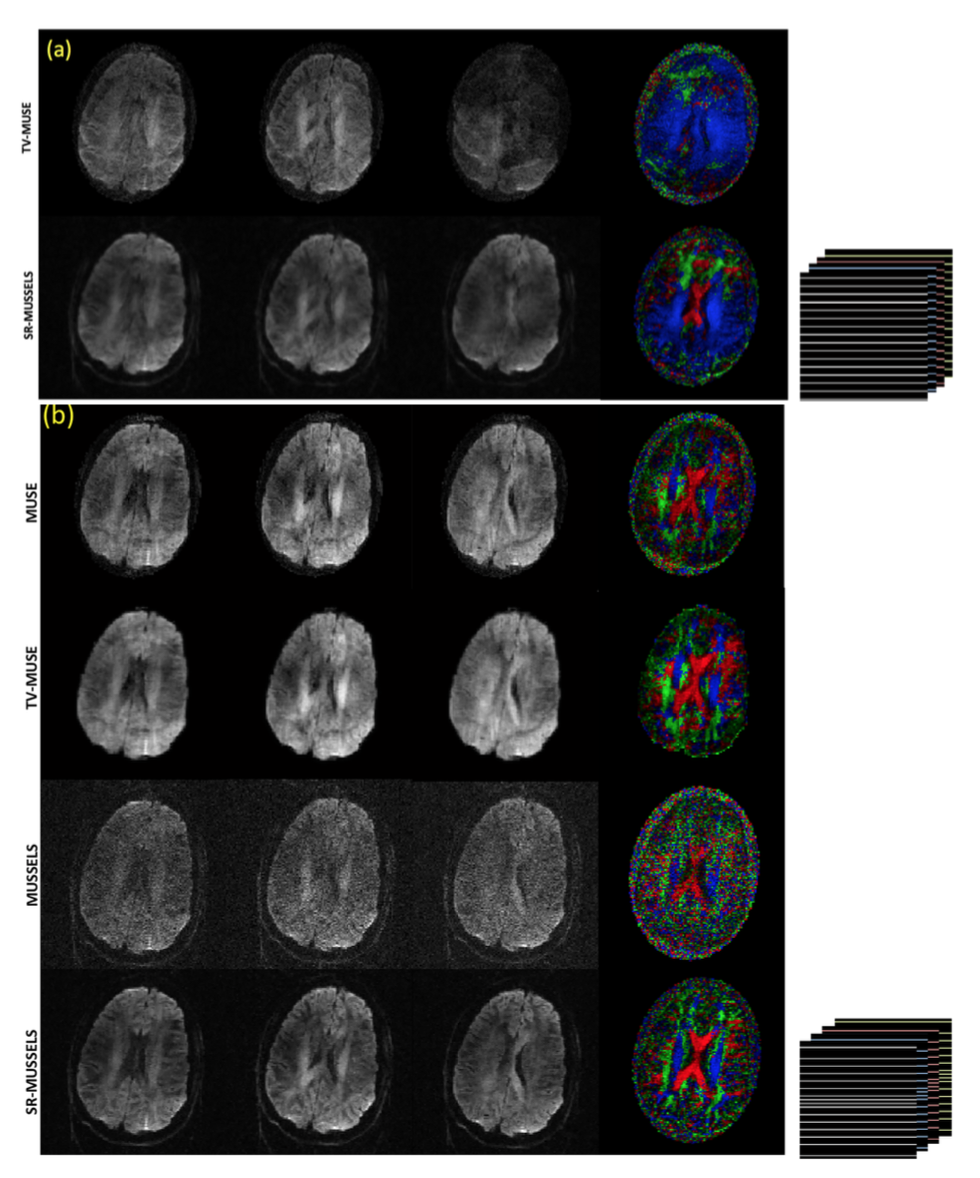}\\
\caption{4 shot under-sampled MS-DW data. (a) shows the TV-MUSE and SR-MUSSELS reconstruction of 3 DWIs reconstructed from uniformly under-sampled data. The under-sampling pattern is shown on the right side. (b) shows the unregularized as well as regularized MUSE and MUSSELS reconstruction of 3 DWIs reconstructed from non-uniformly under-sampled data. The under-sampling pattern used is included which shows that the center-k-space lines were kept intact for all the shots. The unregularized reconstructions also performed reasonably well with the non uniform under-sampling pattern.}
\label{fig:compareus_muse_mussels}
\end{figure}
\clearpage

\newpage

\begin{algorithm}
\floatname{algorithm}{Table 1:}
\renewcommand\thealgorithm{}
\begin{algorithmic}[1]
\STATE{Initialize  $\beta > 0, \gamma$}
\STATE{ Initialize the algorithm by channel combining the measured k-space data to form ${\bf{\hat{m}}}^{(0)}= {\bf \cal{A}}^H{\bf \cal{A}}(\bf{\hat y})$. }
\STATE{ set $n=0$}
\STATE{Repeat}
\STATE{${\bf D}^{(n)}={\bf \bf{F}}({\bf \hat m}^{(n)})$ \text{ where } ${\bf{F}}$ \text{ is the block-Hankel matrix given in Eq. [\ref{jointlifting}]}.\\

Compute the singular value decomposition:  $U\Sigma V^T=SVD( {\bf D}^{(n)})$\\
Perform singular value shrinkage using the rule $\Sigma_k=diag\{(\sigma_i - k)_+\}$ where $\sigma$ are the singular values along the diagonal of $\Sigma$.}\\
\STATE{ Update ${\bf D}^{n+1}={\cal {H}}^{*}(U\Sigma_kV^T))$ where ${\cal {H}}^{*}$ is the inverse mapping of the block-Hankel elements into the multi-shot data matrix. }
\STATE{Update ${\bf{\hat{m}}}^{(n+1)}$ by solving $C_2({\bf \hat m})$ given in Eq. [\ref{AL2}] using CG }\\ 
\STATE{Update Lagrange multipliers: $\mathbf{\gamma}^{(n+1)}=\mathbf{\gamma}^{(n)}-\beta(\bf D-{\bf{F}}(\bf \hat m))$}
\STATE{ set $n=n+1$}
\STATE{ Until stopping criterion is reached}
\caption{Augmented Lagrangian algorithm for solving SR-MUSSELS}
\end{algorithmic}
\end{algorithm} 
\clearpage

\end{document}